\documentclass{mem}
\usepackage{natbib}\usepackage{txfonts}\usepackage{balance}\usepackage{subfigure}
\usepackage{graphicx}
\usepackage[a4paper,breaklinks,dvipdfm]{hyperref}
\idline{83}{1}

\begin{document}

\def\Pdot{\dot{P}}
\def\Pddot{\ddot{P}}

%----------------------------------------------------------------------------------------------------

\title{Multiwavelength Campaign of Observations of AE Aqr}
\subtitle{}

\author{%
C.\, W.\, Mauche,\inst{1} 
M.\, Abada-Simon,\inst{2} 
J.-F.\, Desmurs,\inst{3} 
M.\, J.\, Dulude,\inst{4,5} 
Z.\, Ioannou,\inst{6} 
J.\, D.\, Neill,\inst{7} 
A.\, Price,\inst{8} 
N.\, Sidro,\inst{9} 
W.\, F.\, Welsh,\inst{4} 
\and members of the CBA and AAVSO
}

\offprints{C.\ W.\ Mauche}
  
\institute{%
Lawrence Livermore National Lab., L-473, 7000 East Ave., Livermore, CA 94550, USA
\email{mauche@llnl.gov}
\and
LESIA/CNRS UMR8109, Observatoire de Paris, 92195 Meudon, France
\and
Observatorio Astron—mico Nacional, C/Alfonso XII 3, 28014 Madrid, Spain
\and
Astronomy Department, San Diego State University, San Diego, CA 92182, USA
\and
Space Telescope Science Institute, 3700 San Martin Dr., Baltimore, MD 21218, USA
\and
Physics Dept., University of Crete, P.O.~Box 2208, 710 03 Heraklion, Crete, Greece
\and
California Inst.\ of Technology, 1200 East California Blvd., Pasadena, CA 91125, USA
\and
Amer.\ Assoc.\ of Variable Star Observers, 49 Bay State Rd., Cambridge, MA 02138, USA
\and
Institut de F\'isica d'Altes Energies, Edifici Cn., E-08193, Bellaterra, Spain
}

\authorrunning{Mauche et al.}
\titlerunning{Multiwavelength Campaign of Observations of AE Aqr}

\abstract{We provide a summary of results, obtained from a multiwavelength (TeV $\gamma$-ray, X-ray, UV, optical, and radio) campaign of observations of AE Aqr conducted in 2005 August 28--September 2, on the nature and correlation of the flux variations in the various wavebands, the white dwarf spin evolution, the properties of the X-ray emission region, and the very low upper limits on the TeV $\gamma$-ray flux.

}
\maketitle{}

%----------------------------------------------------------------------------------------------------

\section{Introduction}

AE Aqr is a bright ($V\approx 11$) novalike cataclysmic variable consisting of a magnetic white dwarf primary and a K4--5 V secondary with a long 9.88 hr orbital period and the shortest known 33 s white dwarf rotation period. Although originally classified and interpreted as a disk-accreting DQ Her star, AE Aqr is now widely believed to be a former supersoft X-ray binary \citep{sch02} and 
current magnetic propeller \citep{wyn97}, with most of the mass lost by the secondary being flung out of the binary by the magnetic field of the rapidly rotating white dwarf.

%----------------------------------------------------------------------------------------------------
\begin{figure*}[t!]
\label{Fig1}
\begin{center}
\resizebox{4.25in}{!}{\includegraphics[clip=true]{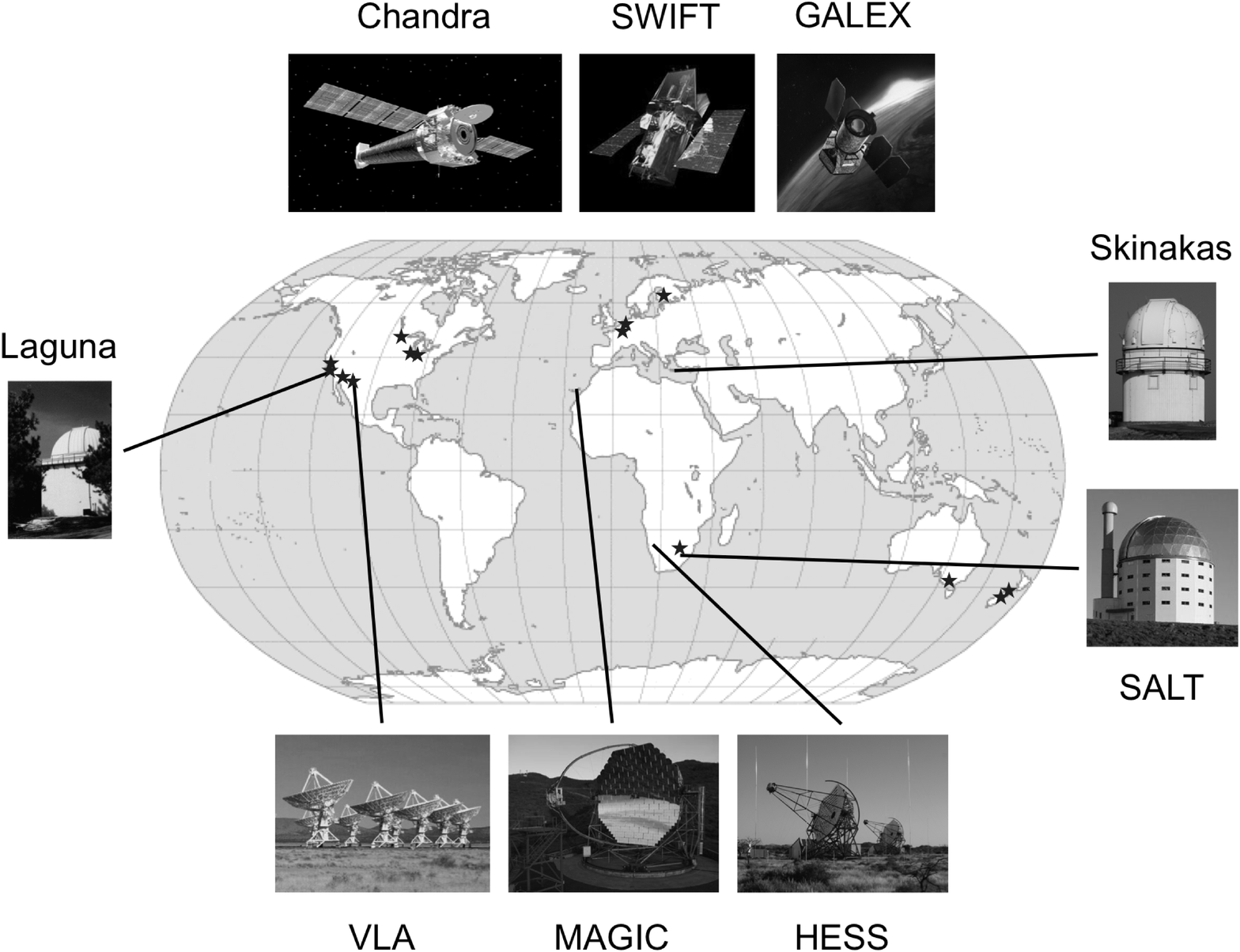}}
\resizebox{4.25in}{!}{\includegraphics[clip=true]{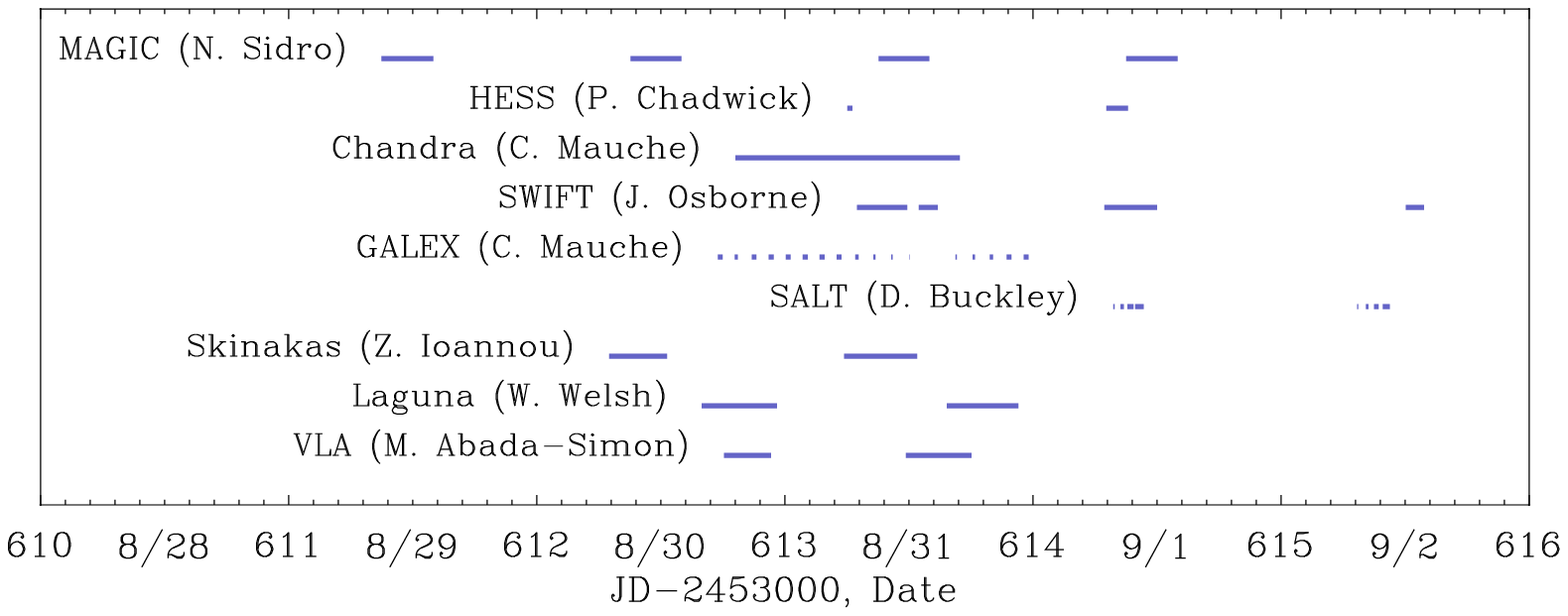}}
\caption{\footnotesize %
{\it Top:\/} Location of facilities participating in the 2005 multiwavelength campaign of observations of AE Aqr; stars mark the locations of CBA and AAVSO astronomers. {\it Bottom:\/} Timeline of the observations, with the name of the lead individual for each facility noted in parentheses.}
\end{center}
%\vskip -10pt
\end{figure*}
%---------------------------------------------------------------------------------------------------- 

Because of its unique properties and variable emission across the electromagnetic spectrum, AE Aqr has been the subject of numerous studies, including the campaign of multiwavelength observations in 1993 October (\citealt{cas96}, and the series of papers in \citealt{buc94}). Given the many improvements in observing capabilities since that time, we undertook a campaign of multiwavelength observations of AE Aqr in 2005 August 28--September 2, built around a {\it Chandra\/}/{\it HST\/}/VLA joint observing proposal. To these coordinated \hbox{X-ray,} UV, and radio observations, we added H.E.S.S.\ and MAGIC TeV $\gamma$-ray observations, Skinakas Observatory 1.3-m telescope and Mount Laguna Observatory 40-in telescope fast B-band optical photometry, and extensive B- and V-band optical photometry obtained by Center for Backyard Astrophysics (CBA) and American Association of Variable Star Observers (AAVSO) professional amateur astronomers B.\ Allen, M.\ Bonnardeau (BZU), P.\ de Ponthiere (DPP), A.\ Gilmore (GAM), K.\ A. Graham (GKA), R.\ A. James (JM), M.\ Koppelman (KMP), B.\ Monard (MLF), A.\ Oksanen (OAR), T.\ Richards (RIX), D.\ Starkey (SDB), and T.\ Vanmunster (VMT) (see Fig.\ 1). 

%----------------------------------------------------------------------------------------------------
\begin{figure*}[t!]
\label{Fig2+3}
\begin{center}
\resizebox{5.01563in}{!}{\includegraphics[clip=true]{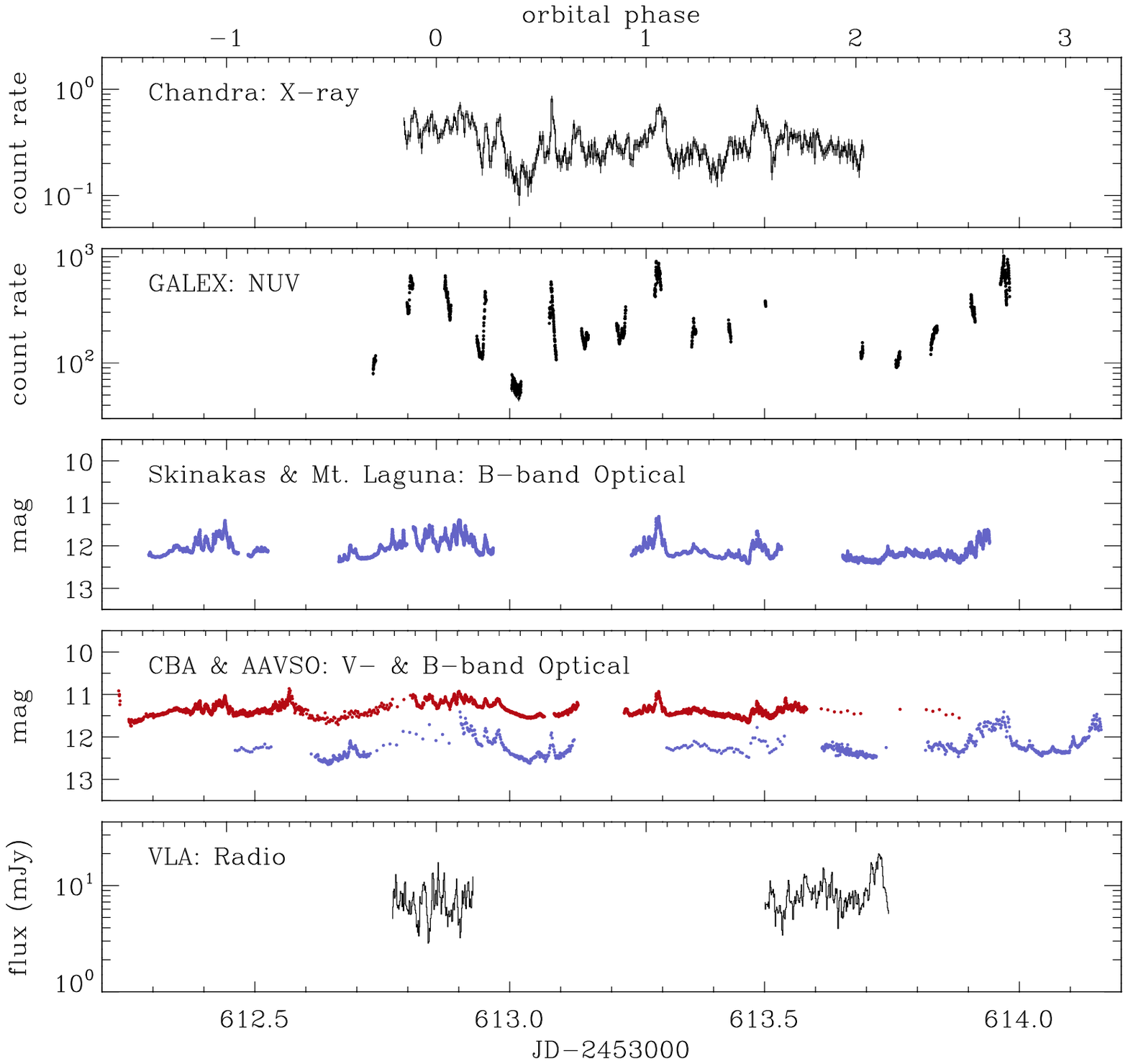}} % (4.5/6.0)*6+11/16
\caption{\footnotesize %
{\it Chandra\/} X-ray; {\it GALEX\/} NUV; Skinakas, Mount Laguna, CBA, and AAVSO B- and V-band ({\it blue and red points, respectively}) optical; and VLA radio light curves of AE Aqr.}
\resizebox{5.01563in}{!}{\includegraphics[clip=true]{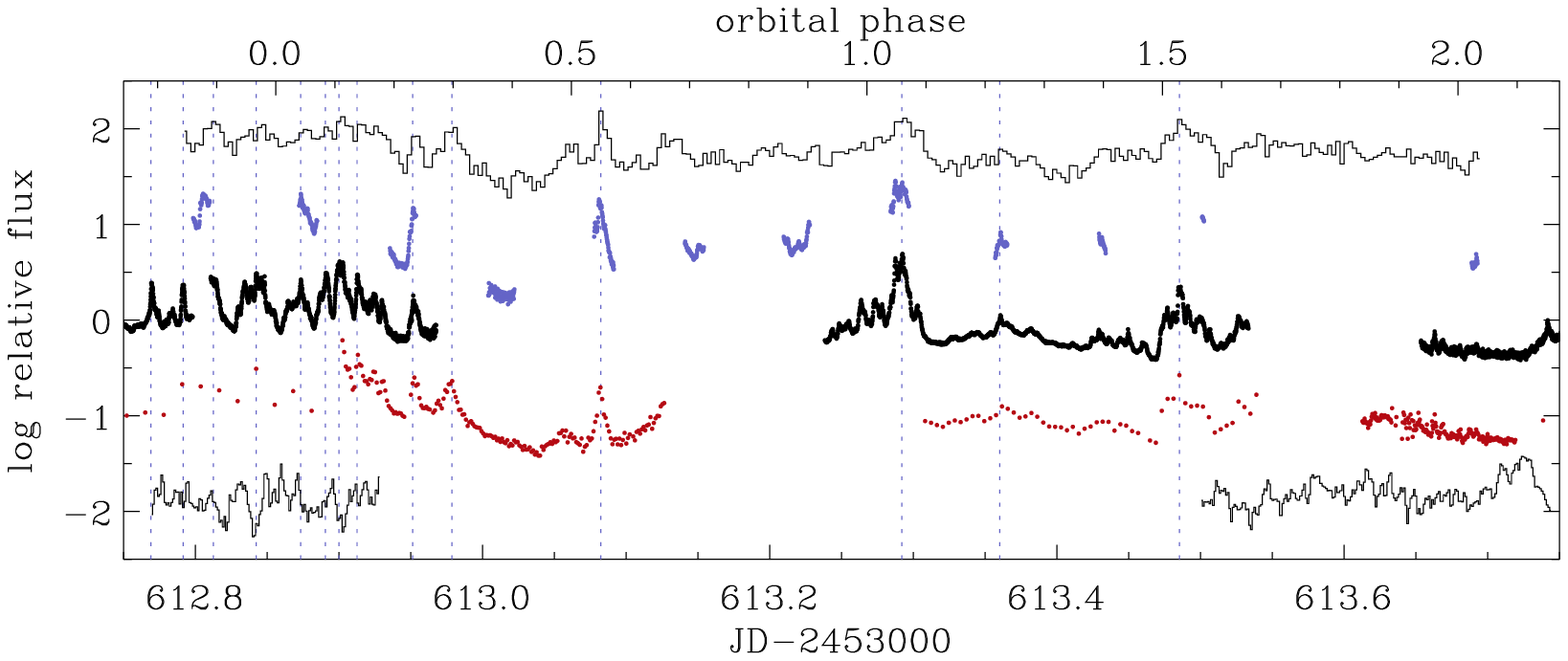}} % (4.5/6.0)*6+11/16
\caption{\footnotesize %
{\it Chandra\/} X-ray ({\it upper histogram\/}); {\it GALEX\/} NUV ({\it blue points\/}); Skinakas \& Mount Laguna ({\it black points\/}) and CBA \& AAVSO ({\it red points\/}) B-band optical; and VLA radio ({\it lower histogram\/}) light curves of AE Aqr. The times of some of the more prominent optical flares are marked with dotted blue vertical lines.}
\end{center}
\vskip -10pt
\end{figure*}
%----------------------------------------------------------------------------------------------------

%----------------------------------------------------------------------------------------------------
\begin{figure*}[t!]
\label{Fig4}
\begin{center}
\resizebox{5.01563in}{!}{\includegraphics[clip=true]{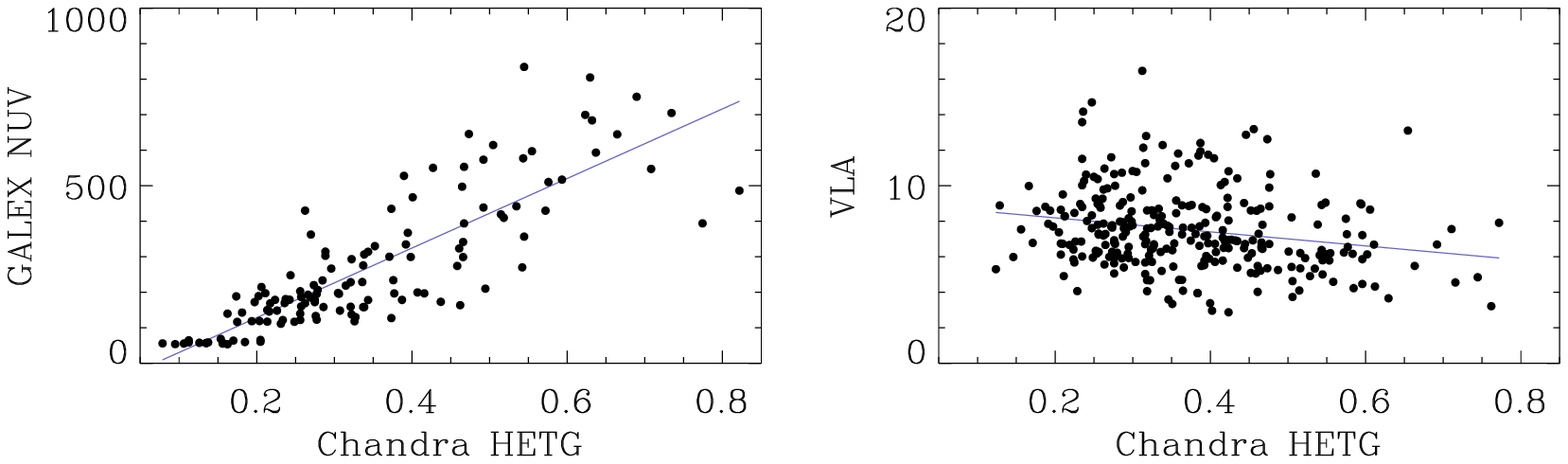}} % (4.5/6.0)*6+11/16
\caption{\footnotesize %
Correlation between the {\it Chandra\/} HETG count rate and the {\it GALEX\/} NUV count rate ({\it left\/}) and the VLA flux ({\it right\/}). Linear fits to the data ({\it blue lines\/}) have correlation coefficients of $+$0.83 and $-$0.22, respectively.}
\end{center}
%\vskip -10pt
\end{figure*}
%----------------------------------------------------------------------------------------------------

As is maddeningly typical of multiwavelength campaigns, our efforts met with mixed success. H.E.S.S.\ was able to observe AE Aqr for only two brief intervals; the weather was less than ideal during the MAGIC observations; our plan to obtain {\it HST\/} STIS time-tagged UV echelle grating spectroscopy was thwarted by the failure of the FUV-MAMA detector's power supply; our backup plan to \hbox{obtain} {\it GALEX\/} FUV and NUV photometry was compromised by a then ongoing problem with the FUV detector; the timings of the pulsations observed in the {\it GALEX\/} NUV data are uncertain because of an uncertain offset to the spacecraft clock; the observations were spread out over a longer interval and had less overlap than is ideal, and some of the acquired data never have been analyzed. Nonetheless, we have published results to summarize and a number of new results to report.

\section{Light Curves}

The {\it Chandra\/} HETG (0.5--6 keV) X-ray, {\it GALEX\/} NUV (1750--2800~\AA ) ultraviolet, Skinakas and Mount Laguna Observatory B-band optical, CBA and AAVSO B- and V-band optical, and VLA 3.6 cm radio light curves obtained during our campaign are shown in Figs.\ 2 and 3. Although the flux is variable in every waveband, the nature of the variability in the radio waveband is clearly different from that in the optical through X-ray wavebands. The flares in these wavebands are highly correlated, last between a few hundred and a few thousand seconds, and have an amplitude that increases from the optical through the NUV. To quantify the degree of correlation between the various wavebands, we show in Fig.\ 4 scatter plots of the {\it GALEX\/} NUV and VLA radio flux as a function of the {\it Chandra\/} HETG X-ray flux. Linear fits to the data, shown by the blue lines, have correlation coefficients of $+$0.83 and $-$0.22, respectively, demonstrating a strong positive correlation between the X-ray and NUV wavebands and a weak negative correlation between the X-ray and radio wavebands. We take these results as evidence that the optical through X-ray emission regions are tightly coupled and hence in close proximity, whereas the radio emission region is not.

%----------------------------------------------------------------------------------------------------
\begin{figure*}[t!]
\label{Fig5}
\begin{center}
\resizebox{5.01563in}{!}{\includegraphics[clip=true]{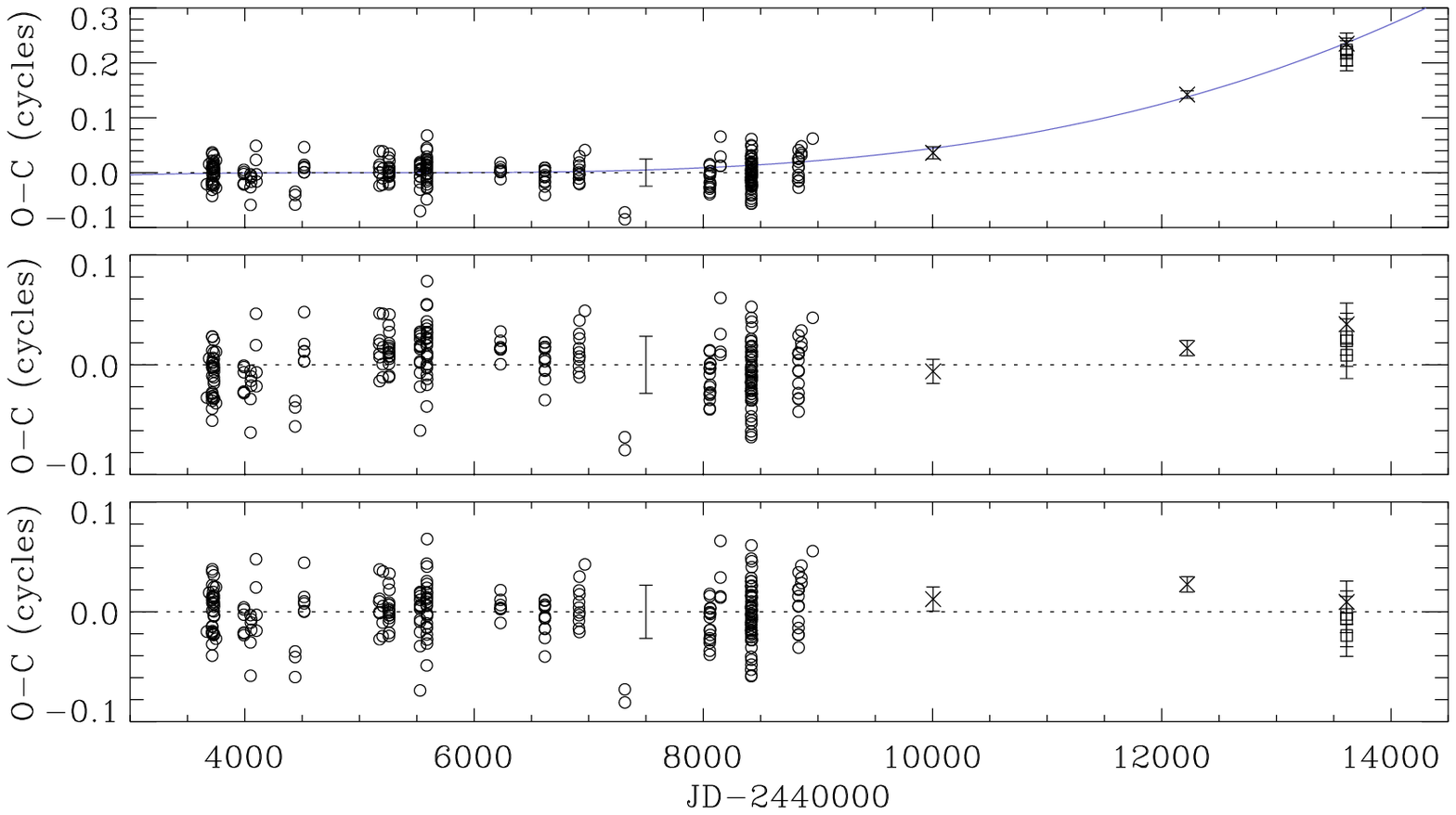}} % (4.5/6.0)*6+11/16
\caption{\footnotesize %
$O-C$ residuals of fits to the optical timings of \citet{deJ94} [{\it circles\/}], the {\it ASCA\/}, {\it XMM-Newton\/}, and {\it Chandra\/} X-ray timings of \citet{mau06} [$\times$s], and the Skinakas and Mount Laguna optical timings of \citet{dul09} [{\it squares\/}] for the \citet{deJ94} quadratic ephemeris ({\it top\/}) and the quadratic ({\it middle\/}) and cubic ({\it bottom\/}) ephemerides given in Table 1. Standard deviations of the data relative to the fits are shown by the error bars. Blue curve in the top panel is the additional $\Pddot= 3.46\times 10^{-19}~{\rm d}^{-1}$ cubic term to the \citet{deJ94} ephemeris proposed by \citet{mau06}.}
\end{center}
%\vskip -10pt
\end{figure*}
%----------------------------------------------------------------------------------------------------

%----------------------------------------------------------------------------------------------------
\begin{table*}[t!]
\caption{Spin Ephemeris Constants}
\vskip -25pt
\label{table}
\begin{center}
\begin{tabular}{lcccc}
\hline
& $T_0$ (BJD) & $P\ ({\rm d})$ & $\Pdot\ ({\rm d~d}^{-1})$ & $\Pddot\ ({\rm d}^{-1})$ \\
\hline
\hbox to 0.5in{de Jager\leaders\hbox to 0.4em{\hss.\hss}\hfill} &
2445172.0000423(10) & 0.00038283263840(28) & $5.642(20)          \times 10^{-14}$ & $\cdots$ \\
\hbox to 0.5in{Quadratic\leaders\hbox to 0.4em{\hss.\hss}\hfill}&
2445172.0000392(12) & 0.00038283263735(18) & $5.752(\phantom{2}8)\times 10^{-14}$ & $\cdots$ \\
\hbox to 0.5in{Cubic\leaders\hbox to 0.4em{\hss.\hss}\hfill}    &
2445172.0000428(11) & 0.00038283263823(17) & $5.599(\phantom{2}7)\times 10^{-14}$ & $4.97(31)\times 10^{-19}$ \\
\hline
\end{tabular}
\end{center}
%\vskip -10pt
\end{table*}
%----------------------------------------------------------------------------------------------------

\section{Spin Period Ephemeris}

\citet{deJ94} used optical and UV photometry of AE Aqr, obtained over the interval 1978 June--1992 November, to show that (1) the 33 s pulsation follows the white dwarf around the binary center of mass, with a projected semi-amplitude $(a\sin i)/c=2.04\pm 0.13$ s, and (2) the rotation period of the white dwarf is spinning down at a rate $\Pdot = 5.642(20)\times 10^{-14}$. \citet{mau06} used the {\it Chandra\/} data obtained during our campaign to show that the 33 s X-ray pulsation also follows the white dwarf around the binary center of mass, with a projected semi-amplitude $(a\sin i)/c=2.17 \pm 0.48$ s. He also found that pulse timings derived from {\it Chandra\/} and from previous {\it XMM-Newton\/} (2001 November) and {\it ASCA\/} (1995 October) observations are not consistent with the de Jager quadratic spin ephemeris, although the ephemeris could be fixed with the addition of a cubic term $\Pddot = 3.46(56)\times 10^{-19}$ ${\rm d}^{-1}$. \cite{dul09} derived pulse timings from three out of four nights of the Skinakas and Mount Laguna B-band optical photometry \hbox{obtained} during our campaign. He found that the optical pulse timings are consistent with the {\it Chandra\/} X-ray pulse time within the errors, and used all the available data to further investigate the white dwarf spin evolution. We discovered a few errors in his transcription of the de Jager pulse timing data, so redid this analysis for both quadratic and cubic ephemerides. The resulting fit residuals are shown in Fig.\ 5 and the spin ephemeris constants are listed in Table 1, which includes the de Jager quadratic spin ephemeris constants for reference.

In each case, the fits are unweighted and the error estimates are derived by assuming that the timing errors are given by the standard deviation of the data relative to the fit (hence, $\chi_\nu^2=1$). While the cubic ephemeris is not required by the data, the spin-down rate $\Pdot$ in the quadratic ephemeris differs from de Jager's value by $5.1 \sigma$; regardless of how it is parameterized, we confirm that the white dwarf in AE Aqr is spinning down at a rate that is faster than predicted by \citet{deJ94}.

%----------------------------------------------------------------------------------------------------
\begin{figure*}[t!]
\label{Fig6}
\begin{center}
\resizebox{5.01563in}{!}{\includegraphics[clip=true]{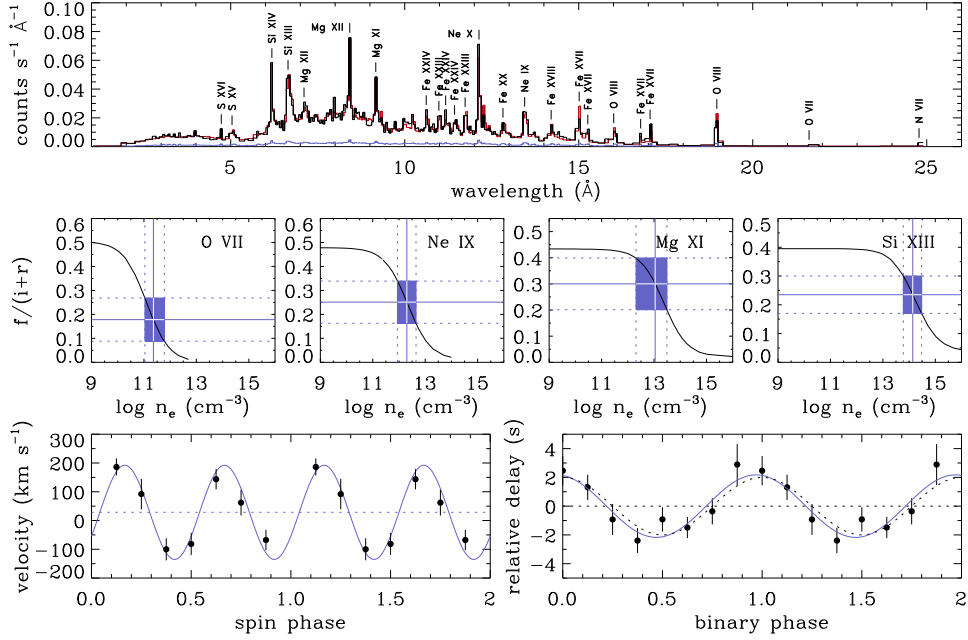}} % (4.5/6.0)*6+11/16
\caption{\footnotesize %
{\it Top:\/} {\it Chandra\/} HETG spectrum of AE Aqr ({\it black histogram\/}), $1\sigma$ error vector ({\it blue histogram\/}), and the best-fit model described in the text ({\it red histogram\/}).
{\it Middle:\/} He-like $f/(i+r)$ line ratios of O~VII, Ne~IX, Mg~XI, and Si~XIII. Blue rectangles delineate the $1\sigma$ error envelope of the measured line ratio and inferred $\log n_{\rm e}$ for each ion.
{\it Bottom left\/}: Spin-phase radial velocities of the X-ray emission lines ({\it filled circles with error bars\/}), best-fitting sine function ({\it solid blue curve\/}), and $\gamma $ velocity ({\it dotted horizontal line\/}).
{\it Bottom right\/}: Orbit-phase delay in the arrival of the X-ray pulse maxima ({\it filled circles with error bars\/}), best-fitting cosine function ({\it solid blue curve\/}), and 2-s semi-amplitude cosine modulation observed in the optical ({\it dotted curve\/}). Adapted from Figs.\ 7, 8, and 10 of \citet{mau09} and Fig.\ 1 of \citet{mau06}.}
\end{center}
%\vskip -10pt
\end{figure*}
%----------------------------------------------------------------------------------------------------

\section{X-ray}

We summarize here and in Fig.\ 6 the results obtained by \citet{mau09} from our {\it Chandra\/} HETG observation of AE Aqr. First, the X-ray spectrum is that of an optically thin multi-temperature thermal plasma; the X-ray emission lines are broad, with widths that increase with the line energy, from $510~\rm km~s^{-1}$ for O VIII to $820~\rm km~s^{-1}$ for Si XIV; the X-ray spectrum is well fit by a plasma model with a Gaussian emission measure distribution that peaks at $\log T ({\rm K})=7.16$, has a width $\sigma=0.48$, an Fe abundance equal to $0.44$ times solar, and other metal (primarily Ne, Mg, and Si) abundances equal to $0.76$ times solar; and for a distance $d=100$ pc, the total emission measure $EM=8\times 10^{53}~\rm cm^{-3}$ and the 0.5--10 keV luminosity $L_{\rm X}= 1\times 10^{31}~\rm erg~s^{-1}$. Second, based on the $f/(i+r)$ flux ratios of the forbidden ($f$), intercombination ($i$), and recombination ($r$) lines of the He $\alpha $ triplets of N VI, O VII, and Ne IX measured by \cite{ito06} in the {\it XMM-Newton\/} RGS spectrum and those of O VII, Ne IX, Mg XI, and Si XIII in the {\it Chandra\/} HETG spectrum, either the plasma electron density increases with temperature by over three orders of magnitude, from $n_{\rm e}\approx 6\times 10^{10}~\rm cm^{-3}$ for N VI [$\log T({\rm K})\approx 6$] to $n_{\rm e}\approx 1\times 10^{14}~\rm cm^{-3}$ for Si XIII [$\log T ({\rm K})\approx 7$], and/or the plasma is significantly affected by photoexcitation. Third, the radial velocity of the X-ray emission lines varies on the white dwarf spin phase, with two oscillations per spin cycle and an amplitude $K\approx 160~\rm km~s^{-1}$.

\section{TeV $\gamma$-ray}

We summarize here the results obtained by \citet{sid08} during four consecutive nights and a total of 15.3 hr of observations of AE Aqr with the 17-m MAGIC atmospheric Cherenkov telescope. Unfortunately, on all four nights of observations, the cosmic trigger rate was low and the optical extinction was high, indicating that the sky was veiled by calima (airborne dust and sand, blown off the Sahara Desert). All of the MAGIC data were analyzed using the standard reconstruction and analysis software, and a set of special Monte Carlo simulations were employed to correct the energy scale and gamma efficiency of our analysis. Because of the relatively poor observing conditions, the lack of a detection, and the absence an entirely independent analysis of the data --- because the analysis has not been thoroughly vetted by the MAGIC consortium --- the following results should be considered preliminary. Nevertheless:

No evidence was found for {\it steady\/} $\gamma $-ray emission on {\it any\/} night or during the {\it sum\/} of the four nights of our observations. On three of the four nights, the upper limits for steady emission are $\sim 8.0\times 10^{-12}~\rm photons~cm^{-2}~s^{-1}$ above 340 GeV and $\sim 2.5\times 10^{-12}~\rm photons~cm^{-2}~s^{-1}$ above 650 GeV. Note that, using a total of 68.7 hr of exposure of AE Aqr obtained with the Whipple Observatory 10-m atmospheric Cherenkov telescope, \citet{lan98} derived an upper limit for steady $\gamma $-ray emission of $4\times 10^{-12}~\rm photons~cm^{-2}~s^{-1}$ above 900 GeV. These upper limits are three orders of magnitude lower than the flux reported by others during the purported $\gamma $-ray burst activity.

In addition to {\it steady\/} emission, we searched for {\it pulsed\/} $\gamma $-ray emission on {\it each\/} night and during the {\it sum\/} of the four nights of our observations, using both the Rayleigh test and phase-folding on frequencies near the white dwarf spin frequency and its first harmonic. No evidence above $2\sigma $ was found for pulsed $\gamma $-ray emission on any of the four nights, with an upper limit of $\sim 1.5\times 10^{-12}~\rm photons~cm^{-2}~s^{-1}$ above 650 GeV.

\section{Summary and Conclusions}

\begin{itemize}
\item
We interpret the strong correlation in the flux variations between the X-ray and NUV wavebands in AE Aqr as evidence that the flaring component of the optical through X-ray emission regions are tightly coupled and hence in close proximity.

\item
We interpret the weak negative correlation in the flux variations between the X-ray and radio wavebands in AE Aqr as evidence that the flaring component of the radio emission region is not tightly coupled to and hence distinct from the optical through X-ray emission regions.

\item
The close association of the optical through X-ray emission regions is demonstrated by the common phase and 2-s semi-amplitude of the binary-phase dependence of the pulse timings. More specifically, the pulsating component of the optical through X-ray emission regions follow the white dwarf around the binary center in mass.

\item
Although the fit parameters of the spin ephemeris of the white dwarf depend on how the pulse timings are modeled and how the pulse timing errors are handled, we confirm that the white dwarf in AE Aqr is spinning down at a rate that is faster than predicted by the quadratic ephemeris of \citet{deJ94}.

\item
The results of the {\it Chandra\/} HETG observation of AE Aqr \citep{mau09}, specifically the high inferred plasma densities and the spin-phase radial velocity variations of the X-ray emission lines, are inconsistent with the recent models of \cite{ito06}, \citet{{ikh06}}, and \citet{ven07} of an extended, low-density source of X-rays in AE Aqr, but instead support earlier models in which the dominant source of X-rays is of high density and in close proximity to the white dwarf.

\item
Given the non-detections of AE Aqr by the Whipple Observatory \citep{lan98} and MAGIC \citep{sid08}
atmospheric Cherenkov telescopes, and absent any publication of a positive TeV $\gamma $-ray detection since the mid-1990's, we suggest the possibility that, contrary to previous claims and common belief, AE Aqr is not a TeV $\gamma $-ray source.
\end{itemize}

%----------------------------------------------------------------------------------------------------
\begin{acknowledgements}
We are grateful to J.\ Patterson and J.\ Kemp of the CBA, A.\ Henden and E.\ Waagen of the AAVSO, and our CBA and AAVSO colleagues B.\ Allen, M.\ Bonnardeau, P.\ de Ponthiere, A.\ Gilmore, K.\ A.\ Graham, R.\ A.\ James, M.\ Koppelman, B.\ Monard, A.\ Oksanen, T.\ Richards, D.\ Starkey, and T.\ Vanmunster for their contributions to this campaign.

Support for this work was provided in part by NASA through (1) {\it Chandra\/} Award Number GO5-6020X issued by the {\it Chandra\/} X-ray Observatory Center, which is operated by the Smithsonian Astrophysical Observatory for and on behalf of NASA under contract NAS8-03060 and (2) Interagency Agreement NNG06EA98I in support of {\it GALEX\/} program ID GALEXGI04-0000-0043; GALEX is operated for NASA by the California Institute of Technology under NASA contract NAS5-98034.

This work was performed under the auspices of the U.S.\ Department of Energy by Lawrence Livermore National Laboratory under Contract DE-AC52-07NA27344.

\end{acknowledgements}
%----------------------------------------------------------------------------------------------------

%----------------------------------------------------------------------------------------------------
\bibliographystyle{aa}

%----------------------------------------------------------------------------------------------------

%----------------------------------------------------------------------------------------------------
\bigskip
\bigskip
\noindent {\bf DISCUSSION}

\bigskip
\noindent {\bf ROBERT CONAN SMITH:} Given the measured radial velocity amplitude of the X-ray emission lines and the known spin period, can you estimate the radius at which the X-ray plasma is trapped?

\bigskip
\noindent {\bf CHRIS MAUCHE:} Evidently not. The measured $K$ velocity of the X-ray emission lines is approximately $160~\rm km~s^{-1}$, whereas the rotation velocity of material trapped on, and rotating with, the white dwarf is $2\pi r/P_{\rm spin}\approx 1300\> (r/R_{\rm wd})~\rm km~s^{-1}$. A simple two-spot model of the X-ray emission region produces $K\approx 500~\rm km~s^{-1}$, but this is still too large by a factor of $\sim 3$. This remains a puzzle.

\bigskip
\noindent {\bf MANABU ISHIDA:} You said that the estimate of the size of the X-ray emission region will be reduced by an order of magnitude relative to that given by \citet{ito06}; isn't it still larger than the size of the white dwarf? Also, the maximum plasma temperature is too low for accretion onto the white dwarf.

\bigskip
\noindent {\bf CHRIS MAUCHE:} I actually said {\it orders\/} of magnitude: At the peak of the emission measure distribution, the emission measure $EM\approx 5.8\times 10^{52}~\rm cm^{-3}$, the density $n_{\rm e}\approx 1\times 10^{14}~\rm cm^{-3}$, and the linear scale $l=(EM/n_{\rm e}^2)^{1/3}\approx 1.4\times 10^8~{\rm cm}\sim 0.2~R_{\rm wd}$. I interpret the observed low maximum plasma temperature in AE Aqr as evidence that the accreting plasma does not pass through a strong hydrodynamic shock on its way to the white dwarf surface. The absence of such a shock might be expected for the low accretion rate, hence the low specific accretion rate, in AE Aqr.

\bigskip
\noindent {\bf PIETER MEITJES:} I disagree with your claim that AE Aqr is not a TeV source. We found in early TeV observations that the duty cycle is very small ($<$10\%). So, many more observations, especially with H.E.S.S., are required.

\bigskip
\noindent {\bf CHRIS MAUCHE:} Although I agree that the question of AE Aqr's purported TeV emission can be settled only with additional observations, Whipple and MAGIC already have observed AE Aqr for a combined 84 hr and seen nothing. Either we have been extraordinarily unlucky, or the original claims that AE Aqr is a TeV source are wrong. I feel lucky.

%----------------------------------------------------------------------------------------------------
\end{document}